\documentclass[fleqn,usenatbib,useAMS]{mnras}

\usepackage{graphicx}	
\usepackage{amsmath}	
\usepackage{amssymb}	

\usepackage{xspace}

\usepackage[dvipsnames]{xcolor}

\newcommand{\pow}[1]{\ifmmode{}^{#1}\else ${}^{#1}$\fi}

\newcommand{\cm}{\,\ifmmode{{\mathrm{cm}}}\else cm\fi}

\newcommand{\ergps}{\,{\rm erg}\,{\rm s}\ifmmode{}^{-1}\else${}^{-1}$\fi}
\newcommand{\Mpch}{\,{\rm Mpc}\,\ifmmode h^{-1}\else $h^{-1}$\fi}
\newcommand{\snru}{\,\ifmmode{\mathrm{Myr}^{-1}}\else Myr${}^{-1}$\fi}
\newcommand{\kms}{\,\ifmmode{\mathrm{km}\,\mathrm{s}^{-1}}\else km\,s${}^{-1}$\fi\xspace}

\newcommand{\cl}{\mathrm{cl}}
\defcitealias{McCourt2016}{M18}

\newcommand{\lshatter}{\ifmmode{\ell_{\mathrm{shatter}}}\else $\ell_{\mathrm{shatter}}$\fi}
\def\gsim{\;\rlap{\lower 2.5pt
 \hbox{$\sim$}}\raise 1.5pt\hbox{$>$}\;}
\def\lsim{\;\rlap{\lower 2.5pt
   \hbox{$\sim$}}\raise 1.5pt\hbox{$<$}\;}

\usepackage[T1]{fontenc}
\usepackage{ae,aecompl}

\title{Is Multiphase Gas Cloudy or Misty?}

\author[M. Gronke \& S. P. Oh]{
  Max Gronke\thanks{E-mail: maxbg@ucsb.edu, Hubble fellow}
  and S. Peng Oh
\\
Department of Physics, University of California, Santa Barbara, CA 93106, USA
}

\date{Accepted 2020 February 11. Received 2020 February 5; in original form 2019 December 17}

\pubyear{2020}

\begin{document}
\label{firstpage}
\pagerange{\pageref{firstpage}--\pageref{lastpage}}
\maketitle

\begin{abstract}
  Cold $T \sim 10^{4}$K gas morphology could span a spectrum ranging from large discrete clouds to a fine `mist' in a hot medium. This has myriad implications, including dynamics and survival, radiative transfer, and resolution requirements for cosmological simulations. Here, we use 3D hydrodynamic simulations to study the pressure-driven fragmentation of cooling gas.
  This is a complex, multi-stage process, with an initial Rayleigh-Taylor unstable contraction phase which seeds perturbations, followed by a rapid, violent expansion leading to the dispersion of small cold gas `droplets' in the vicinity of the gas cloud. Finally, due to turbulent motions, and cooling, these droplets may coagulate.
  Our results show that a gas cloud `shatters' if it is sufficiently perturbed out of pressure balance ($\delta P/P\sim 1$), and has a large \textit{final} overdensity $\chi_{\mathrm{f}}\gtrsim 300$, with only a weak dependence on the cloud size. Otherwise, the droplets reassemble back into larger pieces. We discuss our results in the context of thermal instability, and clouds embedded in a shock heated environment.  
 \end{abstract}




\section{Introduction}
\label{sec:intro}
Understanding the formation and dynamics of cold ($\sim 10^4\,$K) gas is a crucial facet of galaxy formation. Cold gas not only fuels star formation, but (in contrast to the hot phase) it is detectable up to high redshift, and thus widely used as a probe of galactic outflows or the circumgalactic medium \citep[e.g.,][]{Veilleux2005,Tumlinson2017}. 
However, despite its importance and ubiquity, cold gas around galaxies remains an enigma. It is present in galactic halos for a wide range of galaxy masses \citep[e.g.,][]{Steidel2010a,Wisotzki2015}, but its origin, perhaps from thermal instability or ejection from the galaxy, is poorly understood. Such cold gas should be vulnerable to disruption by hydrodynamic instabilities \citep{klein94,zhang17}. 
Furthermore, it has been established observationally
through multiple probes that circumgalactic cold gas can be structured on both small scales ($< 100\,$pc, and potentially substantially less; \citealp[e.g.,][]{Rauch1999,Churchill2003a,Schaye2007,Hennawi2015}), and large scales \citep[$\sim 100$ kpc; e.g.,][]{werk14}. The origin, survival, and morphology of cold gas are all outstanding puzzles.

\begin{figure*}
  \centering
  \includegraphics[width=\textwidth]{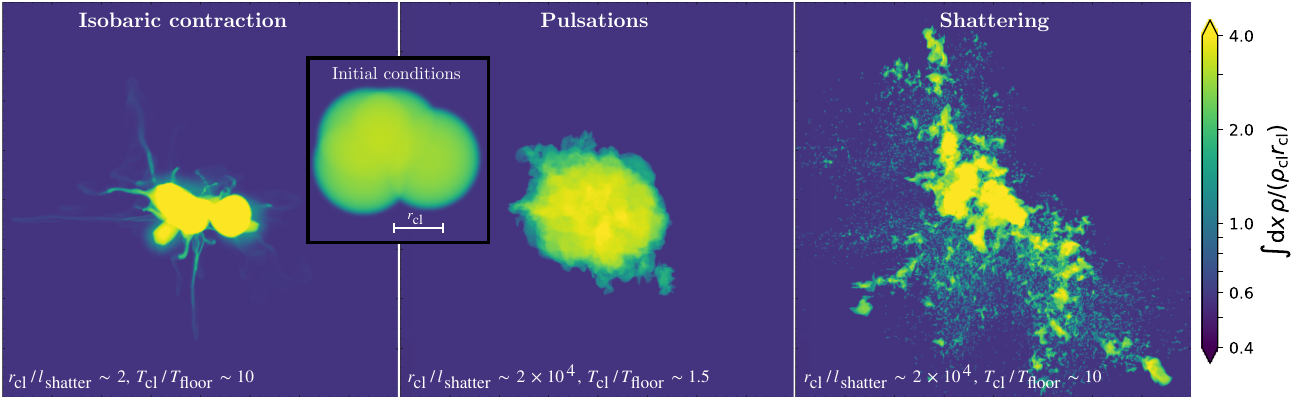}
  \vspace{-1.5em}
  \caption{The outcome of different evolutionary paths of a cooling cloud, for a small cloud ($r_{\rm cl} \sim\lshatter$; left panel), and large clouds ($r_{\rm cl} \gg \lshatter$) which grow mildly (central) and strongly (right) out of pressure balance. Shown are the column densities of runs with $r_{\mathrm{cl}} / l_{\mathrm{cell}}\sim 64$; the parameters are stated in the bottom left corner of each panel, and the snapshots shown are at times (from left to right):
    $t \sim \{1,\, 9400,\,860\}t_{\mathrm{cool,cl}}\sim \{51,\,7,\,5\}t_{\mathrm{sc,floor}}$.
    The inset shows the initial conditions to scale. While the small cloud in the left panel cooled isobarically, the larger clouds either pulsate (central panel) or shatter into tiny pieces (right panel), depending on the degree of pressure imbalance.}
  \vspace{-1em}
  \label{fig:multipanel_options}
\end{figure*}

Observationally, significant evidence for small-scale structure exists (e.g., reviewed in  \citealp{McCourt2016}, henceforth: \citetalias{McCourt2016}), e.g., the order unity area covering fraction of dense gas in the CGM, despite the small volume filling fraction ($f_{\rm V} \lsim 0.1-1\%$). Theoretically, 
\citetalias{McCourt2016} showed that when the cooling time falls far below the sound-crossing time in a cooling cloud, it becomes strongly underpressured relative to surrounding hot gas, 
and `shatters' 
to cloudlets of size
$\lshatter\!\sim\!\min(c_{\rm s} t_{\rm cool})\! \sim\! 0.1 \, {\rm pc} \, (n/{\rm cm^{-3}})^{-1}$, akin to gravitational fragmentation to the Jeans length. Thus, `clouds' of cold, atomic gas may have the structure of a mist, with tiny dispersed fragments embedded in ambient gas.
\citet{Sparre2019} and \citet{Liang2018} subsequently investigated 3D hydro and 2D MHD shattering in a wind-tunnel like setup respectively, while \citet{Waters2019-linear} studied linear non-isobaric thermal instability. 

On the other hand, in \citet{Gronke2018,Gronke2019} \citep[also see][]{2019MNRAS.tmp.3180L}, where we revisited cloud entrainment in a wind, we found that cold gas could survive hydrodynamic instabilities only if clouds {\it exceed} a critical lengthscale $r_{\rm min}  > c_{\rm s} t_{\rm cool,mix} \gg$ \lshatter. This criterion arises from $t_{\mathrm{cool, mix}}< t_{\mathrm{cc}} $, where $t_{\mathrm{cool,mix}}$ is the cooling time of the mixed warm gas and $t_{\mathrm{cc}}$ is the cloud-crushing time. In this regime, the cooling of mixed, `warm' gas causes the cold cloud mass to grow. The cloud retains its monolithic identity; cold gas only survives as large `clouds'. The mass growth rate is similar in nearly static simulations with weak shear. The cloud pulsates due to loss of pressure balance seeded by radiative cooling, entraining hot gas which subsequently cools.  

The `misty' and 'cloudy' scenarios may appear mutually contradictory. However, terrestrially, we experience both; in the ISM and CGM, there is observational evidence for both. What is not known are the physical conditions under which cold gas should exist primarily in a `misty' or `cloudy' state, which we address here. This question has important  consequences for the survival  and dynamics of cold gas, the required resolution for a converged CGM \& IGM in cosmological simulations \citep[][]{VandeVoort2018,Hummels2018,Peeples2018,2019ApJ...881L..20M}, and radiative transfer and escape of ionizing and resonant line photons \citep[e.g.,][]{gronke17}.

\vspace{-2em}
\section{Numerical setup}
\label{sec:setup}

`Shattering' appears to take place when a large ($r_{\rm cl} \gg$ \lshatter)  cloud falls out of pressure balance with its surroundings. This occurs in at least two physical situations: (i) thermal instability, when the cloud pressure falls precipitously due to radiative cooling; (ii) a shock engulfing cold clouds, when the surrounding gas pressure rises sharply. 

To simulate this, we placed four spherical clouds with radius $r_\cl$ and overdensity $\chi_{\rm i}$ inside the simulation domain of size $8r_\cl$ per dimension. We placed one cloud in the center, but displaced the three others with a maximum offset per dimension of $r_{\mathrm{cl}}$ (see inset in Fig.~\ref{fig:multipanel_options} for initial conditions). The deviation from spherical symmetry avoids the carbuncle instability \citep{Moschetta2001}, which seeds grid-aligned artifacts. We varied $\chi_{\mathrm{i}}$ between $1.01$ and $1000$; the lower, (higher) value mimics a linear overdensity such as the origin of thermal instability, and cold gas embedded in a shock heated environment respectively. 
Furthermore, we perturb the density in every cell by a random factor $r$ which is drawn from a Gaussian with $(\mu,\,\sigma) = (1,\,0.01)$ (truncated at $3\sigma$). 

We set the initial cloud temperature to be $T_{\cl} > T_{\mathrm{floor}}$, where the cooling floor is $T_{\mathrm{floor}}=4\times 10^4\,$K and $T_{\cl} / T_{\mathrm{floor}}$ varies from $1.5$ to $1000$. The cloud is initialized to be in pressure balance with its surroundings, but rapidly falls out of pressure balance via radiative cooling. Since the cooling time is much shorter than the sound crossing time, the cloud cools isochorically to $T_{\mathrm{floor}}$ and $P_{\rm cl}/P_{\rm hot} \approx T_{\mathrm{floor}}/T_{\cl} < 1$. Thus, varying $T_{\cl}/T_{\mathrm{floor}}$ is equivalent to varying the degree of initial pressure imbalance; we tested this explicitly. The cloud can only regain pressure balance at a higher overdensity $\chi_{\rm f} = \chi_{\rm i} T_{\cl}/T_{\mathrm{floor}}$. We define \lshatter$\equiv c_{\mathrm{s,floor}} t_{\mathrm{cool}}(\rho_{\mathrm{f}},\, T_{\mathrm{floor}})$.


To emulate heating of the background hot gas, we inhibited cooling for $T > 0.6 T_{\mathrm{hot}}$.
We performed most of our simulations using $r_\cl / l_{\mathrm{cell}} = 16$ cell elements but we increased the resolution to $r_\cl / l_{\mathrm{cell}} = 64$ for some runs as indicated in the text. Note that we do {\it not} resolve \lshatter\ in our simulations; shattering proceeds down to grid scale. While the morphology of `shattered' gas is not numerically converged, we have explicitly checked that our conclusions about the presence/absence of shattering, i.e., the fragmentation of the perturbed gas cloud (the focus of this paper) is numerically robust to resolution. We find  that the resolution requirements to observe shattering are less stringent in 3D than in the 2D simulations of \citetalias{McCourt2016}.

The simulations were run using \texttt{Athena} 4.0 \citep{Stone2008} using the HLLC Riemann solver, second-order reconstruction with slope limiters in the primitive variables, and the van Leer unsplit integrator \citep{Gardiner2008}, and the \citet{Townsend2009} cooling algorithm (using a $7$-piece powerlaw fit to the \citet{Sutherland1993} solar metallicity cooling function). 
The runtime of the simulations was max $(10 t_{\mathrm{sc,cl}}, t_{\mathrm{cool,cl}}, 10 t_{\mathrm{cool,floor}})$, or until the cloud clearly shattered, with a significant number of droplets leaving the simulation domain.
Animations visualizing our numerical results are available at \url{http://max.lyman-alpha.com/shattering}.

\begin{figure}
  \centering
  \includegraphics[width=.95\linewidth]{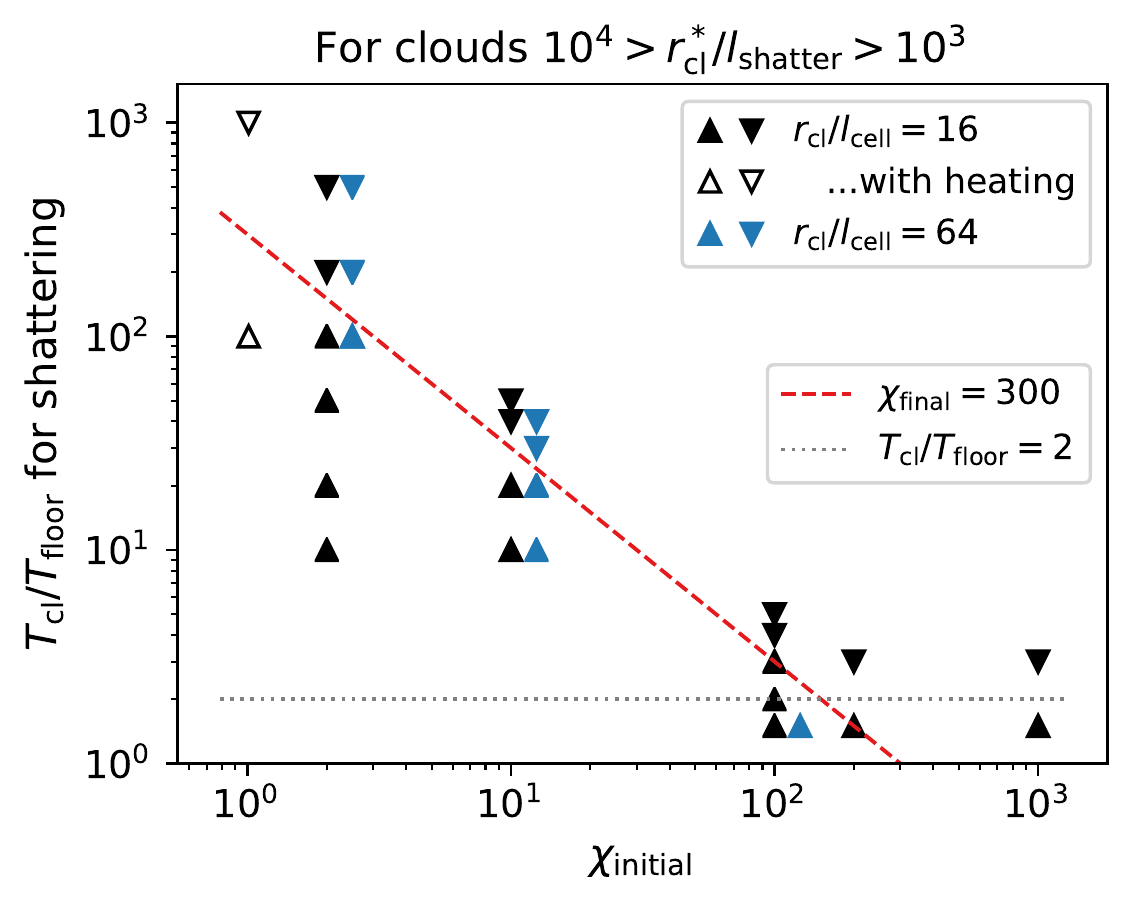}
  \vspace{-1.2em}
  \caption{Conditions required for a large cloud to shatter. The triangles pointing downwards (upwards) show simulations that did (did not) shatter. The dashed line represents the final overdensity $\chi_{\rm f} \approx (T_{\rm cl}/T_{\rm floor}) \chi_{\rm i}=300$, which separates the two regimes. Open symbols represent runs with heating. 
    $r_\cl / \lshatter=64$ are slightly shifted in $x$ direction for clarity.
  }
   \vspace{-2em}
  \label{fig:Xshatter_vs_chi}
\end{figure}

\begin{figure}
  \centering
  \includegraphics[width=\linewidth]{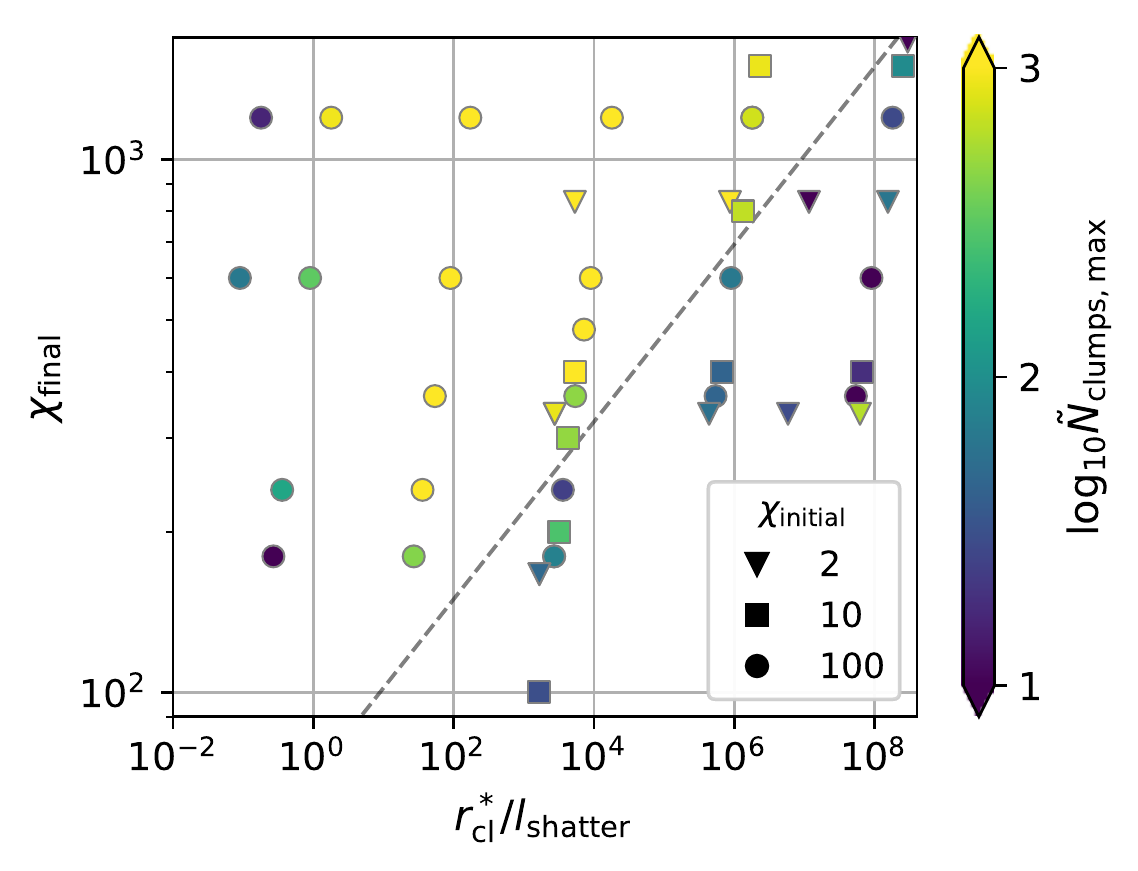}
  \vspace{-1.2em}
  \caption{The resolution-adjusted maximum number of clumps identified during the runtime of a simulation, as a function of cloud size when sonic contact is lost (see text) and final overdensity $\chi_{\rm f}$. The number of clumps shows a sharp transition at $\chi_{\rm crit}$ for `shattering' given by the dashed line and $r_{\cl}\sim l_{\mathrm{shatter}}$ (see text for details). The marker shape indicates the initial overdensity $\chi_{\mathrm{i}}$. Points might be slightly offset for visualization.}
  \label{fig:overview}
  \vspace{-1em}
\end{figure}

\vspace{-1em}
\section{Results}
\label{sec:results}

Our simulations have 3 variables: cloud size, degree of pressure imbalance, and initial overdensity. Fig.~\ref{fig:multipanel_options} first shows graphically how cloud evolution changes with size and pressure contrast. The left panel shows a small cloud of size $r_{\rm cl} \sim \lshatter$ with large pressure imbalance ($T_{\rm cl} / T_{\mathrm{floor}}\sim 10$), while the central and right panels show large clouds ($r_{\rm cl} \gg$ \lshatter) with small and large pressure imbalances respectively ($T_{\rm cl} / T_{\mathrm{floor}}\sim \{1.5,10\}$). Clearly, both a large cloud {\it and} large pressure imbalance is required for shattering: only the simulation shown in the rightmost panel breaks up into small pieces. Interestingly, shattering does not happen during the initial compression by the surroundings, but only during an expansion phase. 
By contrast, the small cloud (left panel) never drastically loses sonic contact and contracts isobarically, while the large cloud with a smaller pressure imbalance (central panel) does not break up but instead oscillates \citep[similar to what can be found in an entrained, growing cold gas cloud in a galactic wind at lower Mach numbers, where pressure variations are small;][]{Gronke2019}. 

What is the effect of the initial overdensity $\chi_{\rm i}$? We first vary the parameters $(\chi_{\rm i}, T_{\cl}/T_{\mathrm{floor}})$, holding cloud size $r_\cl \gg \lshatter$ roughly constant. Fig.~\ref{fig:Xshatter_vs_chi} shows the simulations which did and did not shatter, with triangles pointing downwards and upwards, respectively. Note that here, and below we define a simulation to `shatter' if  it broke apart to more than $100$ pieces. 
Such fragmentation precludes reassembly back into a monolithic entity.
A less overdense cloud needs to be more out of pressure balance (higher $T_{\rm cl}/T_{\rm floor}$) to shatter. The boundary between these regimes has a scaling $T_{\rm cl}/T_{\rm floor} \propto \chi_{\rm i}^{-1}$, collapsing the criterion to a {\it single} parameter: the final overdensity must exceed a critical value $\chi_{\rm f} \sim (T_{\rm cl}/T_{\rm floor}) \chi_{\rm i} > \chi_{\rm crit} \sim 300$ to shatter. This requirement flattens out at large $\chi_{\rm i} > \chi_{\rm crit}$: at least $T_{\mathrm{cl}} / T_{\mathrm{floor}}\gtrsim 2$ (corresponding to $P_{\rm cl} \lesssim 0.5 P_{\rm hot}$) is required. Note that the numerical value of $\chi_{\mathrm{crit}}$ depends on parameters such as the gas metallicity, and $T_{\rm floor}$; however, we expect the governing physics to be  unchanged.

The case of lower initial overdensities ($\chi_{\rm i} \lesssim 10$) requires special care. Since the initial cooling time is long at low densities, the cloud may initially retain sonic contact, that is $r_\cl \lesssim c_{\mathrm{s,cl}}t_{\mathrm{cool}}(T_{\rm cl}, \rho_\cl)$. However, since $t_{\rm cool}$ plummets as it cools and contracts, it will lose sonic contact at some point (since $r_\cl \gg \lshatter\equiv c_{\mathrm{s,floor}}t_{\rm cool}(T_{\rm floor}, \rho_{\mathrm{floor}})$). The scale important for shattering is in fact the radius the cloud has when it loses sonic contact, which is $r^*_\cl= \sqrt{\gamma P_0 / \rho_{*}} t_{\mathrm{cool}}(P_0 / \rho_{*},\,\rho)$ where $P_0$ is the ambient pressure and $\rho_{*} = \rho_{\cl} (r_{\cl}/r^{*}_{\rm cl})^{3}$ comes from mass conservation.
We use $r^*_\cl$ to evaluate cloud size. If the cloud loses sonic contact before contraction (or never does), we set $r^*_\cl$ to $r_\cl$. 

Can shattering occur during thermal instability? This has not been shown to date; \citetalias{McCourt2016} began with non-linear initial conditions, $\chi_{\rm i} \sim 10$. We simulated linear initial overdensities $\chi_{\mathrm{i}}=1.01$ and $T_{\mathrm{cl}} / T_{\mathrm{floor}} = \{10^2,\,10^3\}$, where there is a long period of slow contraction. 
Here, we did not set the cooling function to zero above some temperature but instead introduced constant volumetric heating, set to equal the total cooling rate at each timestep. The cloud shatters, or does not, for $\chi_{\rm f} \approx \{10^3,\,10^2\}$ respectively (see unfilled triangles in Fig.~\ref{fig:Xshatter_vs_chi}), as expected since these bracket $\chi_{\rm crit} \approx 300$.

Since $(T_{\rm cl}/T_{\rm floor}, \chi_{\rm i})$ collapse to the single variable $\chi_{\rm f}$, the entire parameter space can be viewed in the $(\chi_{\rm f}, r_\cl^*/\lshatter)$ plane. In Fig.~\ref{fig:overview} we show how these two variables affect the maximum number of clumps which appear during the simulation. Larger clouds require higher resolution; the larger initial size $\chi_{\rm i}\sim \{2,\,10\}$ cases are run with $r_\cl / l_{\mathrm{cell}}=64$ and only the $\chi_{\rm i} \sim 100$ runs use the fiducial resolution of $r_\cl / l_{\mathrm{cell}}=16$. To account for this, we use the rescaled variable $\tilde N_{\mathrm{clumps,max}} = N_{\mathrm{clumps}} r_\cl / l_{\mathrm{cell}} / 64$.
This shows a sharp transition at
$\chi_{\rm crit} \approx 300 \left( \frac{r^{*}/l_{\rm shatter}}{5000} \right)^{1/6}$
shown by the dashed line in Fig.~\ref{fig:overview}. The exact scaling may change with resolution, but the conclusion that $\chi_{\rm crit}$ scales only weakly with cloud size is robust.  



\begin{figure*}
  \centering
  \includegraphics[width=\textwidth]{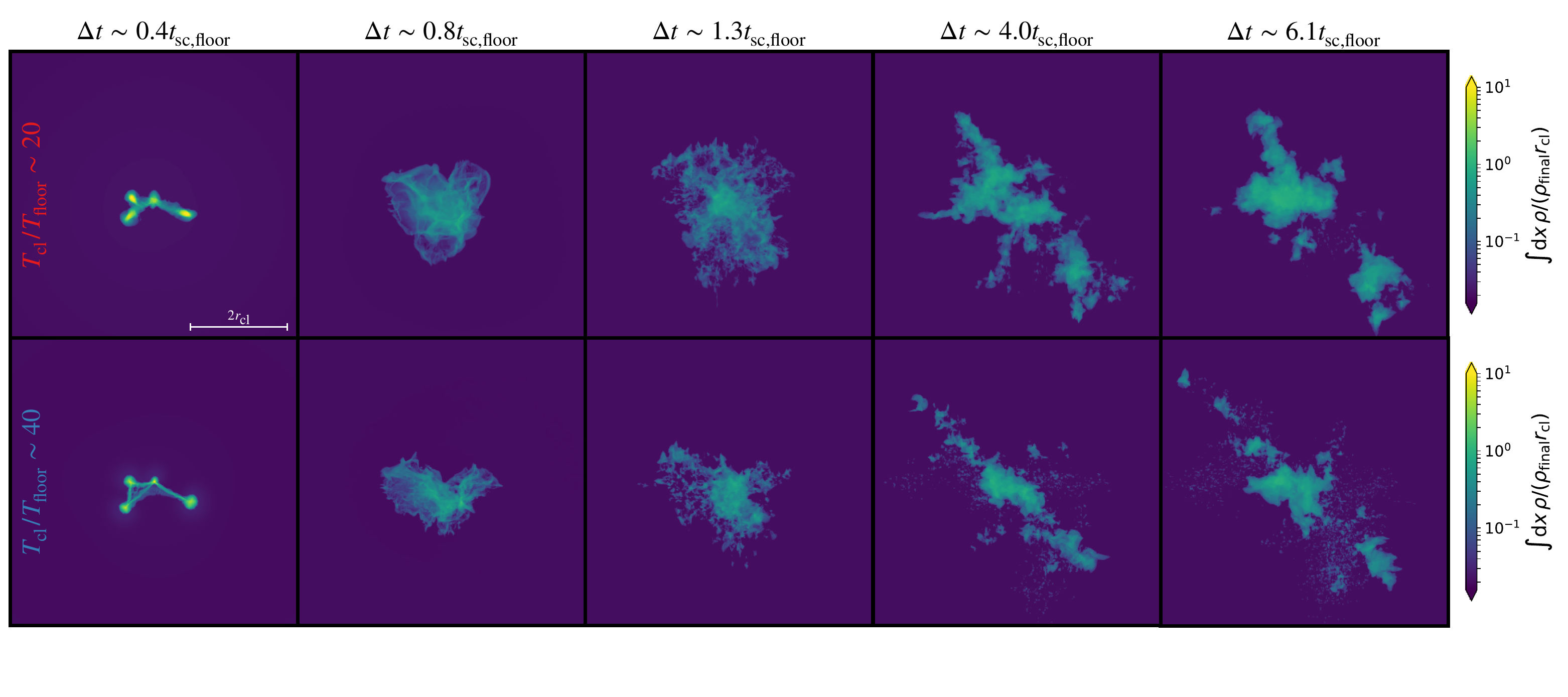}
  \vspace{-1cm}
  \caption{The evolution of shattering. The upper (lower) row shows snapshots from simulations with $\chi_{\mathrm{i}}\sim 10$ and $T_{\cl} / T_{\mathrm{floor}}\sim \{20,40\}$, and thus $\chi_{\rm f} \sim \{200,400\}$ respectively. The time $\Delta t$ is measured relative to the first contraction. Each row shows the process of initial contraction, expansion (with clear fragmentation), shattering (leading to the formation of many ``droplets''), and finally coagulation -- where the droplets merge back again onto the larger cold gas structures. The simulation with the larger $\chi_{\mathrm{final}}$ (shown in the lower row) shows less merging as can also be seen in Fig.~\ref{fig:nclumps_evolution} (where the line color matches the labels in the first panel of this figure).
  }
  \label{fig:shattering_evolution}
\end{figure*}

\begin{figure}
  \centering
  \includegraphics[width=.95\linewidth]{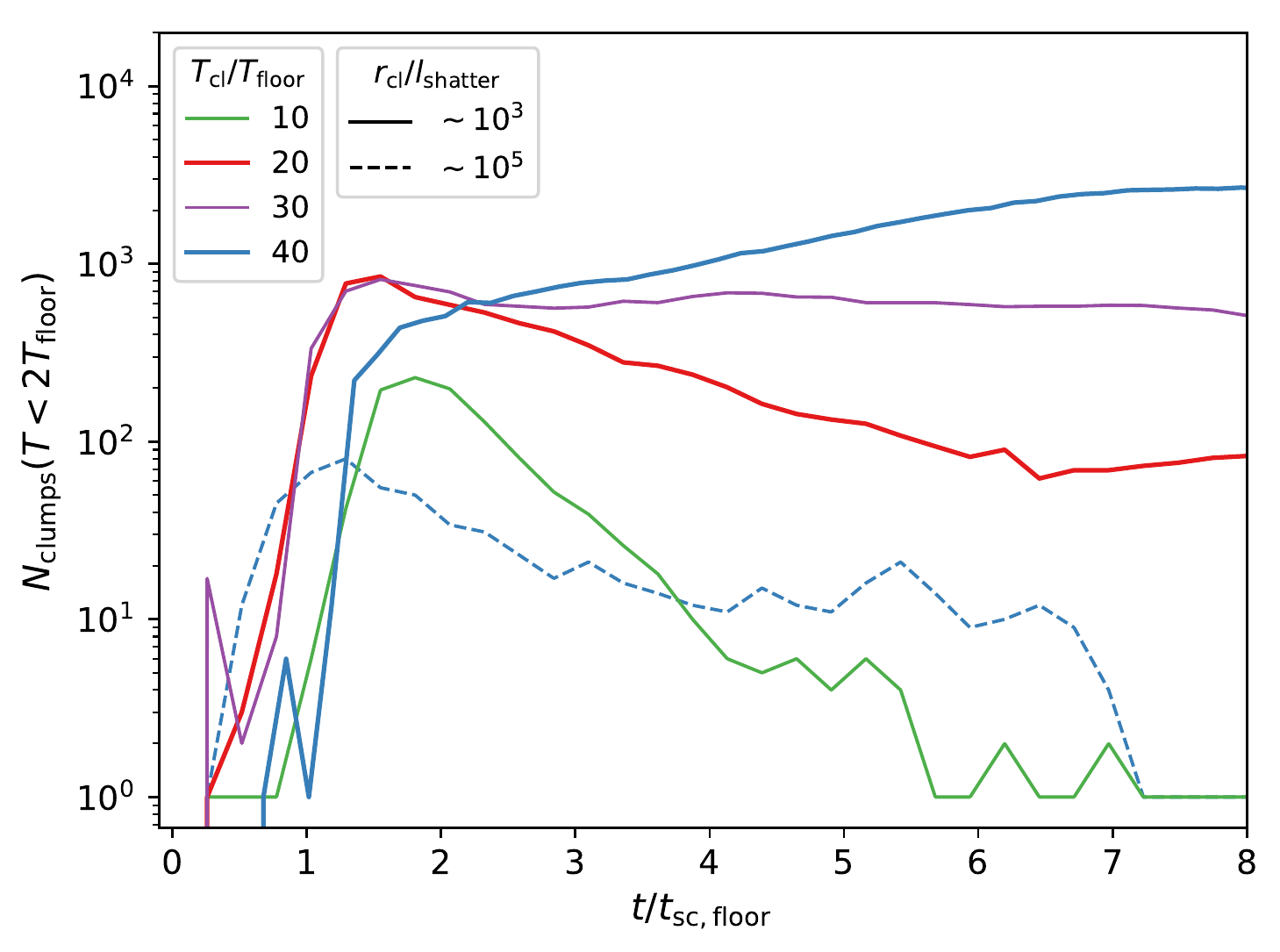}
  \vspace{-1.2em}
  \caption{The number of clumps (with $T < 2 T_{\mathrm{floor}}$) as a function of time for simulations with $\chi_{\mathrm{i}}\sim 10$ but different $\chi_{\mathrm{f}}$ and cloud sizes (denoted by the color and linestyle, respectively). The thicker lines (in red \& blue) are the simulations shown in Fig.~\ref{fig:shattering_evolution}. While all clouds break up into small pieces initially, the efficiency of subsequent coagulation determines the final outcome.}
  \label{fig:nclumps_evolution}
  \vspace{-1em}
\end{figure}

What is the physical origin of shattering? We can gain insight by studying simulations which straddle the $\chi_{\rm crit}$ boundary. Fig.~\ref{fig:shattering_evolution} shows snapshots of two simulations, both with $\chi_{\rm i} \sim 10$ and $r_\cl^*\sim 5\times 10^3 \lshatter$ but differing $T_{\rm cl}/T_{\rm floor}$, such that $\chi_{\mathrm{f}}\sim 200$ and $\sim 400$ (upper and lower row, respectively). The clouds evolve as follows: \textit{(i)} initially, they rapidly cool to the floor temperature.
\textit{(ii)} This leaves a large pressure imbalance leading to cloud contraction (first column). Note that the cloud does {\it not} shatter during the contraction phase, as one might expect. Instead, strong density perturbations and hot gas penetration arise due to Rayleigh Taylor instabilities similar to those which arise in supernovae, except this is an implosion rather than an explosion. The pressure in the cloud overshoots that of the surroundings, leading to \textit{(iii)} a rapid expansion phase (second column). As it expands, the cloud \textit{(iv)} fragments into smaller pieces (`shatters'; third column). This shattering is likely driven by Richtmyer-Meshkov instabilities as the expansion front sweeps over the strong density inhomogeneities created during the contraction phase, creating strong vorticity and breaking up the cloud. We tested this by allowing a similarly overpressurized but uniform cloud to rapidly expand; in this case shattering does {\it not} occur. 

Crucially, the cloud evolution now diverges. As seen in the fourth and fifth columns of Fig.~\ref{fig:shattering_evolution}, the cloudlets can either \textit{(v; a)}  disperse in the surrounding of the original cloud (bottom panels), or \textit{(v; b)} fall back and coagulate into larger clouds (top panels). The remaining (or re-forming) larger clouds \textit{(vi)} continue to pulsate. This process is accompanied by a significant mass growth as entrained hot gas cools, as in \citet{Gronke2018,Gronke2019}. For instance, the $T_{\rm cl} / T_{\mathrm{floor}}\sim 20$ ($\sim 40$) cloud depicted in Fig.~\ref{fig:shattering_evolution} has roughly tripled (doubled) its mass at $t\sim 6 t_{\mathrm{sc,floor}}$.

Thus, while `shattering' always begins in a large cloud out of pressure balance, whether it prevails depends on a competition between breakup and coagulation. This can be quantitatively seen in Fig.~\ref{fig:nclumps_evolution} where we show the number of cloudlets as a function of time. The solid lines depict $r_{\rm cl}/l_{\rm shatter} \sim 10^{3}$ (for which $\chi_{\rm crit} \sim 300$), and $\chi_{\rm f}=\{1,2,3,4\} \times 100$. These all show fragmentation into $\gtrsim 100$ pieces at $t\sim 2 t_{\mathrm{sc,floor}}$, the point of maximum expansion.
However, for $\chi_{\mathrm{f}} < \chi_{\mathrm{crit}}$, the droplets coagulate, reversing the shattering process. 
The dashed line shows that coagulation is more efficient for a larger cloud (i.e. $\chi_{\rm crit}$ increases). 

\vspace{-2em}
\section{Discussion \& Conclusions}
\label{sec:discussion}
In this work we revisit the `shattering' mechanism first identified by \citetalias{McCourt2016} in 2D simulations; we now do so in 3D. Our simulations are governed by 3 dimensionless parameters: the initial cloud overdensity $\chi_{\rm i}$, the initial cloud size $r_{\rm cl}/\lshatter$, and the pressure contrast $P_{\rm cl}/P_{\rm hot} \approx T_{\rm floor}/T_{\rm cl}$. As anticipated, we find that a cloud must be large ($r_{\rm cl}/\lshatter\gg1$) and out of pressure balance ($P_{\rm cl}/P_{\rm hot} \lesssim 0.5$) to shatter. However, we also find an unexpected requirement that the final overdensity $\chi_{\rm f} \gsim \chi_{\rm crit} \approx 300$, with a weak scaling on cloud size. 
Otherwise, the fragments quickly merge. For $T_{\rm floor} \sim 10^{4}\,$K, this suggests that shattering is inefficient in gas with $T_{\rm hot} < 3 \times 10^{6}\,$K. We also show that shattering occurs during linear thermal instability $\delta \rho/\rho \ll 1$. It has hitherto only been shown for large, non-linear overdensities.  


We find `shattering' is really a spectrum spanned by the dispersal of cloudlets and their merger.
On the ends of this spectrum lie clouds which  `explode', violently launching the droplets in all directions, and clouds where the restoring coagulation force is so strong that they never break up but instead pulsate. In between, the fate of cold gas depends on the competition between breakup and coagulation, which is governed by the final overdensity $\chi_{\rm f}$. Coagulation was already seen in the original 2D simulations of \citetalias{McCourt2016} (see their fig. 3, third panel from top; fig. 6, and associated discussion), but its true importance is only now apparent. 

What drives coagulation? There are at least two causes: (i) radiative cooling, which drives pressure gradients in mixed interstitial gas, the source of mass growth discussed in \citet{Gronke2018,Gronke2019}. Cooling-induced coalescence has also been highlighted in \citet{elphick91,Waters2019}, though merger velocities seen in those low overdensity ($\chi\sim$ few) calculations are much smaller than seen there, requiring $\gsim 10^{3}-10^{4}$ cooling times for coalescence, and could not compete with breakup. (ii) Turbulence. It is well known that clumping instabilities driven by particle inertia or wave-particle resonance operate in dust-gas interactions \citep[e.g.,][]{lambrechts16, squire18}, and may also operate in cloudlet-gas interactions. 

Without a quantitative understanding of coagulation, we cannot derive a criterion for shattering. Nonetheless, it seems physically reasonable that the competition between `launching' and `drag' depends on overdensity $\chi_{\rm f}$ (which sets the particle stopping length). For coagulation driven by cooling, we can make the following heuristic argument. Suppose that the dispersion of cloudlets is set by the RM instability. The hot gas punches through pressure gradients with a characteristic velocity $c_{\rm s,hot}$, and cold clouds disperse with a characteristic velocity $\sim \alpha c_{\rm s,hot}$, where $\alpha$ encodes imperfect entrainment. Over a cloud oscillation time, the cloudlets disperse over a volume $l_{\rm launch}^{3} \sim (\alpha c_{\rm s,hot} t_{\rm sc,floor})^{3}$. On the other hand, the volume of interstitial hot gas which is consumed by the cloudlets over this time is $V_{\mathrm{hot}}\sim \dot m t_{\mathrm{sc,floor}} / \rho_{\mathrm{hot}} \sim r_{\mathrm{cl}}^3 (r_{\mathrm{cl}}/\lshatter)^{1/4}$, where we have used the mass entrainment rate $\dot m \sim r_\cl^2 \rho_{\mathrm{hot}} v_{\mathrm{mix}}$ where $v_{\mathrm{mix}}\sim c_{\mathrm{s,floor}}(t_{\rm cool,floor}/t_{\rm sc,floor})^{-1/4}$ \citep{Ji2018,Gronke2019}. For the cloudlets to disperse faster than interstitial hot gas can be consumed, we require $l_{\mathrm{launch}}^3 \gtrsim V_{\mathrm{hot}}$. Using $c_{\rm s,hot}/c_{\rm s,floor} \sim \chi_{\rm f}^{1/2}$ yields $\chi_{\mathrm{f}} \gsim \alpha^{-2} \left(r_\cl/\lshatter \right)^{1/6}$,  
which fits our results for $\alpha \sim 0.12$ (cf. dashed line in Fig.~\ref{fig:overview}). 

Besides thermal instability, our setup can also mimic cold clouds (with large $\chi_{\rm i} \gsim 100$) engulfed by a shock, where the  background pressure rises rapidly.  
Using the shock jump conditions, one can relate $\chi_{\mathrm{crit}}\sim \mathrm{max}(300,\,3\chi_{\mathrm{initial}})$ to a requirement for the Mach number of the wind in order for the cloud to shatter:
$  \mathcal{M}^2 \gtrsim \left[(\gamma+1) \chi_{\mathrm{crit}}/\chi_{\mathrm{i}} + (\gamma - 1) \right]/(2 \gamma)\;$,
which corresponds to $\mathcal{M}\sim 1.6$ for $\chi_{\mathrm{i}}\gtrsim
100$. This is roughly consistent with our wind-tunnel setup in \citet{Gronke2019} where
our fiducial $\mathcal{M}=1.5$ setup was numerically converged, while the higher Mach-number
runs (with $\mathcal{M}=3$ and $\mathcal{M}=6$) were
not.
We attributed this to the larger
compression of the cloud, but breakup via shattering can also drive resolution requirements. In the $\mathcal{M}\sim 1.5$ runs, a solid `tail' behind the cloud forms quickly, while the tail in $\mathcal{M}=6$ simulation is much more diffuse and transient. 

Regardless, the competition between breakup and coagulation will differ when there are background gas motions. It is not at all clear that clouds subject to a wind can both shatter {\it and} survive; the pieces need to coagulate into larger fragments ($> c_{\rm s} t_{\rm cool, mix}$) to survive \citepalias[e.g., as in fig. 6 of][]{McCourt2016}. 
It may well be that while clouds can entrain and grow in transonic winds, they do not survive higher Mach number winds, where shattering into small fragments dominates. The impact of background turbulence is also unclear: it could drive fragmentation, or clumping and coagulation (as for dust). These issues require high resolution simulations with careful attention to convergence. Our physical understanding of the shattering/coagulation mechanisms, and their interaction with extrinsic turbulence, magnetic fields, and cosmic rays, remain tenuous. It will be the subject of future work.



\vspace{-2em}
\section*{Acknowledgments}
\vspace{-.5em}
We thank Yan-fei Jiang, Drummond Fielding, Brent Tan and particularly  Dongwook Lee for discussions on numerical issues, and the referee for helpful comments.
This research made use of \texttt{yt} \citep{2011ApJS..192....9T}.
We acknowledge support from NASA grants NNX17AK58G \& HST-HF2-51409, HST grant HST-AR-15039.003-A, and XSEDE grant TG-AST180036. 
MG thanks JHU for hospitality.
\vspace{-2em}

\bibliographystyle{mnras}
\bibliography{refs}


\bsp	
\label{lastpage}
\end{document}